\documentclass[a4paper]{article}
\usepackage{amsmath,amssymb,amsfonts,bm,xcolor}

\usepackage{algorithm,algorithmic,url}
\newcommand{\algrule}[1][.2pt]{\par\vskip.5\baselineskip\hrule height #1\par\vskip.5\baselineskip}

\newcommand{\dn}{\textrm{d}}       
\newcommand{\dd}{\, \textrm{d}}       

\usepackage{enumitem,hyperref,natbib,multirow,rotating,sectsty,booktabs}           
\allsectionsfont{\sffamily}
\usepackage[margin=1in]{geometry}

\usepackage{setspace}

\title{Accelerating pseudo-marginal Metropolis-Hastings\\ by correlating auxiliary variables}

\author{Johan Dahlin, Fredrik Lindsten, Joel Kronander and Thomas B.\ Sch\"{o}n%
\thanks{E-mail to corresponding author: \textit{johan.dahlin@liu.se}. JD is with the Department of Electrical Engineering, Link{\"o}ping University, Sweden. FL is with the Engineering, University of Cambridge, United Kingdom. JK is with the Department of Science and Technology, Link{\"o}ping University, Sweden. TS is with the Department of Information Technology, Uppsala University, Sweden.}%
}

\begin{document}
\maketitle

\doublespacing
\begin{abstract}
\noindent Pseudo-marginal Metropolis-Hastings (pmMH) is a powerful method for Bayesian inference in models where the posterior distribution is analytical intractable or computationally costly to evaluate directly. It operates by introducing additional auxiliary variables into the model and form an extended target distribution, which then can be evaluated point-wise. In many cases, the standard Metropolis-Hastings is then applied to sample from the extended target and the sought posterior can be obtained by marginalisation. However, in some implementations this approach suffers from poor mixing as the auxiliary variables are sampled from an independent proposal. We propose a modification to the pmMH algorithm in which a Crank-Nicolson (CN) proposal is used instead. This results in that we introduce a positive correlation in the auxiliary variables. We investigate how to tune the CN proposal and its impact on the mixing of the resulting pmMH sampler. The conclusion is that the proposed modification can have a beneficial effect on both the mixing of the Markov chain and the computational cost for each iteration of the pmMH algorithm. \\ 

\noindent This work was conducted independently from the recent arXiv pre-print by \cite{DeligiannidisDoucetPittKohn2015}.\\

\end{abstract}

\newpage
\section{Introduction}
\label{sec:intro}

We are interested in approximating some probability distribution $\bar{\pi}(\theta, u)$ using simulation methods based on Markov chain Monte Carlo (MCMC; \citealp{RobertCasella2004}). In Bayesian inference, $\bar{\pi}(\theta, u)$ could represent the posterior distribution of the parameters $\theta \subset \Theta$ and some auxiliary variables $u \subset \mathcal{U}$, which e.g.\ can be latent states in the model or missing data. Often, we are interested in the marginal \textit{posterior distribution of the parameters} given by 
\begin{align}
	\pi_{\theta}(\theta) 
	= 
	\frac{ p( \theta ) p( y | \theta ) }{ \displaystyle \int_{\Theta} \! p( \theta ) p( y | \theta ) \dn \theta }
	=
	\frac{ p( \theta ) p_{\theta}(y) }{ p(y) },
	\label{eq:intro:parameterposterior}
\end{align}
where $p_{\theta}(y) \triangleq p( y | \theta )$ denotes the \textit{likelihood function} and $p( \theta )$ denotes the \textit{parameter prior distribution}. The denominator $p(y)$ is usually referred to as the \textit{marginal likelihood} or the model evidence. 

In principle, we can apply MCMC methods to sample directly from \eqref{eq:intro:parameterposterior} using e.g.\ the Metropolis-Hastings (MH; \citealp{Metropolis1953,Hastings1970}) algorithm. However, in practice $\pi_{\theta}(\theta)$ can be analytically intractable or too costly from a computationally perspective to evaluate point-wise. By introducing the auxiliary variables $u$, we can sometimes mitigate these problems and obtain an algorithm which can be of practical use. 

This approach is discussed by \cite{AndrieuRoberts2009}, where the pseudo-marginal MH (pmMH) algorithm is introduced to sample from $\bar{\pi}(\theta, u)$ using a standard MH sampler. This approach can be seen as an \textit{exact approximation} of the ideal algorithm, where $\pi_{\theta}(\theta)$ can be recovered by marginalisation. A concrete example is the particle MH (PMH; \citealp{AndrieuDoucetHolenstein2010}), where the auxiliary variables are all the random variables generated in a run of a sequential Monte Carlo (SMC; \citealp{DelMoralDoucetJasra2006}) algorithm. In this setting, the SMC algorithm is used to provide an unbiased estimate of the likelihood function. This is useful for inference in latent variables models such as state space models (SSMs; \citealp{FluryShephard2011,PittSilvaGiordaniKohn2012}), mixture models \citep{FearnheadMeligkotsidou2015} and generalised linear mixed models \citep{TranScharthPittKohn2014}.

A typical problem in pmMH is that the resulting Markov chain can be highly autocorrelated, which corresponds to a sticky chain and poor exploration of the posterior. This is the result of that occasionally we obtain an estimate of the posterior which is much larger than its true value. This problem can typically be mitigated by increasing the number of auxiliary variables $N_u$, which increase the accuracy of the estimate of the target. However, this increases the computational cost for each iteration of the algorithm, which can limit practical use of the pmMH algorithm for many interesting but challenging models. The trade-off between increasing $N_u$ and the number of iterations $K$ in the pmMH algorithm is studied by \cite{PittSilvaGiordaniKohn2012}, \cite{SherlockThieryRobetsRosenthal2015} and \cite{DoucetPittKohn2015}. They conclude that $N_u$ should be selected such that the standard deviation in the log-target estimate is roughly between $1$~and~$2$.

The main contribution of this paper is to introduce a positive correlation in the auxiliary variables $u$ between two consecutive iterations of the pmMH algorithm. Intuitively, this allows us to decrease $N_u$ and therefore the computational cost while keeping the mixing of the Markov chain constant. The correlation is introduced by changing the proposal for $u$ from the standard independent proposal to a random walk proposal in the auxiliary space $\mathcal{U}$. 

The idea for this originates in the field of computer graphics, where a similar approach is used for rendering imagesby sampling valid light paths connecting the light sources in the scene to the camera using MCMC methods. In \cite{KelemenSzirmayAntalCsonka2002}, the authors propose the primary sample path space Metropolis light transport algorithm that operates directly on a set of uniform variates used in a second step to construct a light ray through the scene, using sequential importance sampling. Recently, \cite{HachisukaKaplanyanDachsbacher2014} extended the original algorithm  to include multiple importance sampling (MIS; \citealp{OwenZhou2000,VeachGuibas1997,KronanderSchon2014}) to improve the efficiency further.

The idea of introducing correlation in $u$ between iterations has previously been suggested (independently) in the original particle MCMC (PMCMC) paper by \cite{AndrieuDoucetHolenstein2010} and in the connected discussions by \cite{LeeHolmes2010}. Furthermore, a similar idea of introducing correlation in $u$ is proposed by \cite{AndrieuDoucetLee2012}. The main difference in our contribution is the use of the Crank-Nicolson (CN; \citealp{BeskosRobertsStuartVoss2008,CotterRobertsStuartWhite2013,HairerStuartVollmer2014}) proposal to update $u$ at every iteration of the algorithm. This in contrast with the standard pmMH proposal for $u$, which samples new random variables independently at each iteration of the algorithm. The CN proposal is a natural choice as it is known to scale independently with the dimension of the space. Furthermore, we provide the reader with both a theoretical and numerical analysis of how the mixing and the computational cost is affected and propose an adaptive algorithm based on the theoretical results. \cite{DeligiannidisDoucetPittKohn2015} also proposes (independently) to make use of the CN proposal in this setting.

We consider two different numerical examples to illustrate the properties of the proposed modifications to the pmMH algorithm. In the first example, we infer the parameters in IID data from the Gaussian distribution to analyse how to tune the CN proposal and show how it improves upon the idea proposed by \cite{LeeHolmes2010}. In the second example, we conduct inference of the log-volatility using a stochastic volatility (SV) model with leverage on real-world stock index data.

We continue this paper by introducing the proposed modifications in Section~\ref{sec:inducingcorrelation} and analyse its theoretical properties, in a simplified setting, in Section~\ref{sec:theory}. We end the paper by some numerical illustrations of the proposed method in Section~\ref{sec:results} from which we draw some general conclusions in Section~\ref{sec:conclusions}.

\section{Introducing correlation into the auxiliary variables}
\label{sec:inducingcorrelation}
We would like to sample from the parameter posterior $\pi_{\theta}(\theta)$ in \eqref{eq:intro:parameterposterior}. However, this is not possible using a direct implementation of the MH algorithm as we cannot evaluate $\pi_{\theta}(\theta)$ point-wise. Instead, we introduce the auxiliary variables $u=(u_1,\ldots,u_{N_u})$ as $N_u$ independent standard Gaussian random variables, i.e.\ $u_i \sim \mathcal{N}(0,1)$. We make use of the setup used by \cite{AndrieuRoberts2009} and \cite{DoucetPittKohn2015}. Furthermore, we make no notational distinction between a random variable and its realisation for brevity.

Let the \textit{potential function} $\Phi_{\theta}(u): \Theta \times \mathbb{R}^{N_u} \rightarrow \mathbb{R}$ be defined such that
\begin{align*}
	\mathbb{E} \Big[
	\exp 
	\big( 
	- \Phi_{\theta}(u) 
	\big) 
	\Big]
	=
	c \, \pi_{\theta}(\theta),
	\qquad \qquad
	\forall \,
	\theta \in \Theta,
\end{align*}
and some constant $c>0$. Hence, we can define the \textit{extended target distribution} as
\begin{align}
	\bar{\pi}(\theta,u)
	=
	\frac
	{ \exp \big( - \Phi_{\theta}(u) \big) }
	{c}
	\,
	\mathcal{N}
	\big( u; 0, \mathbf{I}_{N_u} \big),
	\label{eq:inducingcorrelation:extendedtarget}
\end{align}
where $\mathcal{N}(u;0,\mathbf{I}_{N_u})$ denotes a $N_u$-dimensional multivariate standard Gaussian distribution. Note that the parameter posterior is the marginal of $\bar{\pi}(\theta,u)$ as
\begin{align*}
	\int_{\mathbb{R}^{N_u}}
	\!
	\bar{\pi}(\theta,u)
	\dd u
	=
	\frac{1}{c}
	\mathbb{E} \Big[
	\exp \big(
	- \Phi_{\theta}(u) 
	\big)
	\Big]
	= 
	\pi_{\theta}(\theta),
\end{align*}
as required in the exact approximation scheme used in the pmMH algorithm. Hence, we conclude that this is a valid \textit{extended target} for sampling from the parameter posterior $\pi_{\theta}(\theta)$. A concrete example of a suitable potential function follows from the definition in \eqref{eq:intro:parameterposterior}. In this case, we have
\begin{align*}
	\Phi_{\theta}(u)
	&=
	-
	\Big(
	\log \widehat{p}_{\theta}(y; u )
	+
	\log p(\theta)
	\Big), \\
	\exp
	\big(
	-
	\Phi_{\theta}(u)
	\big)
	&=
	\widehat{p}_{\theta}(y; u) 
	p(\theta),
	\\
	\mathbb{E}
	\Big[
	\exp 
	\big(
	- 
	\Phi_{\theta}(u)
	\big)
	\Big]
	&=
	p_{\theta}(y) p(\theta)
	=
	\underbrace{p(y)}_{=c} \pi_{\theta}(\theta),
\end{align*}
where the constant $c$ corresponds to the marginal likelihood and $\widehat{p}_{\theta}(y; u)$ denotes the non-negative and unbiased estimator of the likelihood constructed by the auxiliary variables. (The typical application of the pmMH algorithm makes use of importance sampling or particle filtering to construct such an estimator.)

From \eqref{eq:inducingcorrelation:extendedtarget}, we have that the target is high-dimensional in $u$ whenever $N_u$ is large, and that the extended target correspond to a change of measure from the Gaussian prior. In this setting, we know that the CN proposal is a suitable choice as it is dimension independent \citep{CotterRobertsStuartWhite2013}. An intuition for this can be given by realising that the CN proposal is a discretisation of the Ornstein-Uhlenbeck stochastic differential equation, see e.g.\ \cite{KloedenPlaten1992}. The main difference between using the CN proposal and a standard random walk proposal is its autoregressive nature, which results in the mean-reverting behaviour. As a consequence, the CN proposal has the standard Gaussian distribution as its limiting distribution, which suits us well in this setting. Therefore, we assume a proposal distribution for $u$ and $\theta$ with the form
\begin{align}
	q_{\theta,u} \big( \{\theta',u'\}|\{\theta,u\} \big)
	&=
	q_{\theta} \big( \theta' | \{\theta,u\} \big)
	q_{u} \big( u' | u \big)
	\nonumber \\
	&=
	\mathcal{N} \big( \theta'; \mu(\theta,u), \Sigma(\theta,u) \big) \, \,
	\mathcal{N} \big( u';      \sqrt{1-\sigma_u^2} u, \sigma^2_u \mathbf{I}_{N_u} \big).
	\label{eq:inducingcorrelation:proposal}
\end{align}
This corresponds to using a standard Gaussian random walk for $\theta$, where $\mu(\theta,u)$ and $\Sigma(\theta,u)$ denote a mean and covariance function possibly depending on the auxiliary variables in the previous step, respectively. Note that by introducing $u$ into $q_{\theta}$, we can make use of gradient and Hessian information as proposed by \cite{DahlinLindstenSchon2015a} to improve the performance of the algorithm. The CN proposal for $u$ is parametrised by the  step length $\sigma_u \in (0,1)$, which is determined by the user. We return to discussing how to tune $\sigma_u$ in Section~\ref{sec:theory}. Note that \eqref{eq:inducingcorrelation:proposal} is in contrast with the standard independent pmMH proposal, which we recover for the choice $\sigma_u = 1$.

The corresponding pmMH algorithm for sampling from \eqref{eq:inducingcorrelation:extendedtarget} follows directly from the choice of proposal in \eqref{eq:inducingcorrelation:proposal}. During iteration $k$ of the algorithm, we first sample from \eqref{eq:inducingcorrelation:proposal} to obtain the \textit{candidate parameter} $\{\theta',u'\}$ given $\{\theta_{k-1},u_{k-1}\}$. We then compute the acceptance probability by
\begin{align}
	\alpha \big( \{\theta',u'\}, \{\theta_{k-1},u_{k-1}\} \big)
	=
	1
	\wedge
	\exp \Big( \Phi_{\theta_{k-1}}(u_{k-1}) - \Phi_{\theta'}(u') \Big)
	\frac
	{ q_{\theta} \big( \theta_{k-1} | \{\theta',u' \} \big) }
	{ q_{\theta} \big( \theta'      | \{\theta_{k-1},u_{k-1} \} \big) },
	\label{eq:inducingcorrelation:aprob}
\end{align}
where we make use of the notation $a \wedge b \triangleq \min \{a,b\}$. This follows directly from the properties of the CN proposal \citep{CotterRobertsStuartWhite2013}. In the last step, we accept the candidate parameter, i.e.\ set $\{\theta_k,u_k\} \leftarrow \{\theta',u'\}$, with the probability given by \eqref{eq:inducingcorrelation:aprob}. Otherwise, we reject the candidate parameter and set $\{\theta_k,u_k\} \leftarrow \{\theta_{k-1},u_{k-1}\}$.

The resulting pmMH is presented in Algorithm~\ref{alg:pmmh}. In Line~4, we need to evaluate the potential function in $\{\theta',u'\}$. In this paper, we make use of importance sampling and particle filtering in Section~\ref{sec:results} to construct $\Phi_{\theta}(u)$. Note that Algorithm~\ref{alg:pmmh} only differs from a standard pmMH algorithm in Lines~3 and 4, where we add a CN proposal for $u$ and evaluate the potential function $\Phi_{\theta}(u)$. Hence, implementing the new algorithm only requires minor changes to the existing code.

\begin{algorithm}[!t]
\caption{\textsf{Pseudo-marginal Metropolis-Hastings (pmMH)}}
\textsc{Inputs:} $K>0$ (no.\ MCMC steps), $\theta_0$ (initial parameters), $\Phi_{\theta}(u)$ (potential) and $q_{\theta,u}$ (proposal). \\
\textsc{Output:} $\{\{\theta_1,u_1\}\ldots,\{\theta_K,u_K\}\}$ (approximate samples from $\bar{\pi}(\theta,u)$).
\algrule[.4pt]
\begin{algorithmic}[1]
	\STATE Generate $u_0 \sim p(u_0)$ and compute $\Phi_{\theta_0}(u_0)$.
	\FOR{$k=1$ to $K$}
		\STATE Sample $\{\theta',u'\}$ using the proposal in \eqref{eq:inducingcorrelation:proposal}.
		\STATE Compute $\exp( -\Phi_{\theta'}(u') )$ and $\alpha(\{\theta',u'\},\{\theta_{k-1},u_{k-1} \})$ given by \eqref{eq:inducingcorrelation:aprob}.
		\STATE Sample $\omega_k$ uniformly over $[0,1]$.
		\IF{$\omega_k \leq \alpha(\{\theta',u'\},\{\theta_{k-1},u_{k-1} \})$}
			\STATE Accept $\{\theta',u'\}$, i.e.\
			$\{\theta_k, u_k\} \leftarrow \{\theta',u'\}$. 
		\ELSE
			\STATE 
			Reject $\{\theta',u'\}$, i.e.\
			$\{\theta_k, u_k\} \leftarrow \{\theta_{k-1},u_{k-1}\}$.
		\ENDIF
	\ENDFOR
\end{algorithmic}
\label{alg:pmmh}
\end{algorithm}

In many models, we are required to make use of non-Gaussian random variables to generate samples from the proposal distributions in e.g.\ importance sampling and particle filtering algorithms. Furthermore, we require uniform random variables for the resampling step in the particle filter, see e.g.\ \cite{DoucetJohansen2011}. These variables can be obtained by using a inverse cumulative distribution function (CDF) transformation also known as a quantile transformation. Firstly, we compute a uniform random variable by $u^{\star}=\Phi(u)$, where $\Phi(u)$ denotes the CDF of the standard Gaussian distribution. Seconly, we compute $\bar{u}$ by a quantile transform of $u^{\star}$ using the inverse CDF for the distribution that we would like to simulate a random variable from. There exists other approaches e.g.\ accept-reject sampling that also can be useful for simulating random variables when the CDF cannot be inverted in closed-form. See e.g.\ \cite{Ross2012} or \cite{RobertCasella2004} for more information.

\section{Theoretical analysis}
\label{sec:theory}
The aim of the theoretical analysis in this section is to give some guidance of how to tune $\sigma_u$ to obtain a reasonable performance in the pmMH algorithm. More specifically, we are interested in determining the value of $\sigma_u$ that maximises the mixing and to determine how this translates into a suitable acceptance rate in the algorithm. We begin by setting up the model for the analysis in Section~\ref{sec:theory:setup}, which depends on the multivariate random variable $u$. We then reformulate the model to replace $u$ with a univariate random variable $z$. This enables use to make use of an analysis of a discretised model \citep{GelmanRobertsGilks1996,YangRodriguez2013} in Section~\ref{sec:theory:discretisation} to tune the CN proposal to optimise the mixing in the Markov chain. In Section~\ref{sec:theory:results}, we report the results of this analysis, which we return to In Section~\ref{sec:results} where we compare with the optimal tuning of $\sigma_u$ in some practical scenarios.

\subsection{Setting up the model}
\label{sec:theory:setup}
To accomplish the aforementioned aim we will study the algorithm in a simplified setting. Firstly, since we are primarily interested in the effect of the correlation in the $u$-variables, we will start out by making the simplifying assumption that we fix $\theta$ in the extended target $\bar{\pi}(\theta,u)$. Hence, we would like to analyse sampling from
\begin{align}
  \label{eq:theory:bar-u}
	\bar{\pi}(u) = \frac{ \exp \big( - \Phi(u) \big) }{c} \mathcal{N} \big( u; 0, \mathbf{I}_{N_u} \big),
\end{align}
using the CN proposal $q_u(u'|u)$ in \eqref{eq:inducingcorrelation:proposal}. We believe that this is a reasonable proxy for the complete model \eqref{eq:inducingcorrelation:extendedtarget} since in many cases only local moves are made in the $\theta$ variable. Following \cite{PittSilvaGiordaniKohn2012} and \cite{DoucetPittKohn2015} we also assume that $\Phi(u)$ (which depend on $N_u$) follows a central limit theorem (CLT) and that it therefore can be accurately approximated as being Gaussian distributed:
\begin{align}
  \label{eq:theory:phi-is-normal-1}
  - \Phi(u) &\sim \mathcal{N} \left( - \frac{\sigma_{\Phi}^2}{2}, \sigma_{\Phi}^2 \right),
  & \text{when } u &\sim \mathcal{N} \big( u; 0, \mathbf{I}_{N_u} \big).
\end{align}
for some $\sigma_{\Phi} > 0$.
This assumption follows from the properties of the log-likelihood estimator based on (sequential) importance sampling; see \citealp{PittSilvaGiordaniKohn2012,DoucetPittKohn2015} for further details. This case is of interested to us as we make use of this type of estimator in Section~\ref{sec:results} for computing an estimate of the parameter posterior distribution.

We can readily check that assumption \eqref{eq:theory:phi-is-normal-1} implies that
\begin{align}
  - \Phi(u) &\sim \mathcal{N} \left( \frac{\sigma_{\Phi}^2}{2}, \sigma_{\Phi}^2 \right),
  & \text{when } u &\sim \bar\pi(u),
\end{align}
where $\bar\pi$ is the $N_u$-dimensional distribution defined in \eqref{eq:theory:bar-u}. Consequently, if we define the random variable 
\begin{align*}
z \triangleq \frac{\sigma_\Phi}{2} - \frac{1}{\sigma_\Phi} \Phi(u)
\end{align*}
it follows that the law of $z$ under $\bar\pi$ is given by
\begin{align*}
  \widetilde\pi(z) = \mathcal{N}(z ; \sigma_\Phi, 1).
\end{align*}
Note that $z$ is a one-dimensional variable, which means that we have reduced the high-dimensional target \eqref{eq:theory:bar-u} to one of its one-dimensional marginals. We will study the properties of the CN proposal in this one-dimensional marginal space.

Indeed, assume that $u\sim\bar\pi(u)$ is distributed according to the target \eqref{eq:theory:bar-u} (i.e., it can be viewed as the state of the Markov chain at stationarity) and that $u'|u \sim q_u(u'|u)$ is simulated from the CN proposal in \eqref{eq:inducingcorrelation:proposal}. Define $z$ and $z'$ as above, as transformations of $u$ and $u'$, respectively. From a bivariate CLT we can then obtain a bivariate Gaussian approximation for the vector $(z,z')^{\mathsf{T}}$, suggesting that the CN proposal for $u$ corresponds to the one-dimensional proposal:

\begin{align}
	\widetilde{q}(z'|z) = \mathcal{N} \Big( z'; \sqrt{1 - \sigma^2_z}, \sigma^2_z \Big),
	\label{eq:theoretical:peskunproposal}
\end{align}
for $z$, for some $\sigma_z > 0$. Note that $\sigma_z$ in general differs from $\sigma_u$ and that it depends on the nonlinear transformation $\Phi(u)$. However, it is clear that $\sigma_u = 0 \Rightarrow \sigma_z = 0$ and $\sigma_u = 1 \Rightarrow \sigma_z = 1$ and, intuitively, $\sigma_z$ is a monotone function of $\sigma_u$. We investigate the dependence between these two variables numerically in Section~\ref{sec:results:iid}.

Note that the resulting acceptance probability, for the reduced one-dimensional MH algorithm with target $\widetilde\pi(z)$ and proposal $\widetilde q(z'|z)$, is given by:
\begin{align}
	\widetilde{\alpha}(z,z') = 1 \wedge \exp(\sigma_\Phi z' - \sigma_\Phi z).
	\label{eq:theoretical:peskunaprob}
\end{align}

\subsection{Analysis by discretisation of the state space}
\label{sec:theory:discretisation}
We follow \cite{GelmanRobertsGilks1996} and \cite{YangRodriguez2013} and analyse the mixing of the Markov chain using a state space discretisation approach. We know that the expectation of any integrable \textit{test function} $\varphi: \mathcal{Z} \rightarrow \mathbb{R}$ under the target distribution 
\begin{align*}
	\widetilde{\pi}[\varphi]
	=
	\mathbb{E}_{\widetilde{\pi}}[\varphi(z)]
	=
	\int_{\mathcal{Z}} \!
	\varphi(z)
	\widetilde{\pi}(z)
	\dd z,
\end{align*}
can be approximated by the ergodic theorem \citep{MeynTweedie2009,Tierney1994} as
\begin{align*}
	\widehat{\widetilde{\pi}}[\varphi]
	=
	\frac{1}{K}
	\sum_{k=1}^K
	\varphi \big( z_k \big),
\end{align*}
using the samples $\{z_1,z_2,\ldots,z_K\} \triangleq z_{1:K}$ obtained from the pmMH algorithm. Under the assumption of geometric ergodicity, we have that the error of the estimate obeys a CLT \citep{MeynTweedie2009,Tierney1994} given by
\begin{align*}
	\sqrt{K}
	\Big(
	\widetilde{\pi}[\varphi]
	-
	\widehat{\widetilde{\pi}}[\varphi]
	\Big)
	\stackrel{d}{\longrightarrow}
	\mathcal{N}(0, \nu ),
\end{align*}
where $\nu$ denotes the asymptotic variance,
\begin{align}
	\nu = \mathbb{V}_{\widetilde{\pi}}[\varphi] 
	\cdot
	\underbrace{\left[ 1 + 2 \sum_{\tau=1}^{ \infty } \widehat{\rho}_{\tau} (z) \right]}_{\triangleq \textsf{IACT}( z )},
	\label{eq:theory:avariance}
\end{align}
where $\mathbb{V}_{\widetilde{\pi}}[\varphi]$ and $\textsf{IACT}( z )$ denote the variance of $\varphi(z)$ over $\widetilde{\pi}(z)$ and the integrated autocorrelation time (IACT), respectively. Here, $\rho_{\tau}(z) = \mathsf{corr}(\varphi(z_k),\varphi(z_{k+\tau}))$ denotes the lag-$\tau$ autocorrelation of $\varphi$.

To optimise the mixing, we would like to determine $\sigma_z$ such that $1/\nu$ is as large as possible. To estimate the asymptotic variance, we partition the interval $(z_{\min},z_{\max})$ into $L$ bins of equal length $\Delta = (z_{\max}-z_{\min})/L$. Hence, the center point of bin $l$ is given by $z_l = z_{\min} + ( l - 0.5 ) \Delta$. We can then define the transition matrix $P=\{p_{lm}\}$ following \cite{YangRodriguez2013} by
\begin{align}
	p_{lm} 
	=
	\begin{cases}
		\widetilde{q}(z_m|z_l) \widetilde{\alpha}(z_l,z_m) \Delta, & \text{for } l,m \in \{1,\ldots,K\}, l \neq m \\
		1 - \sum_{k \neq l} p_{lk}, & \text{for } l = m
	\end{cases}
	\label{eq:theory:transitionmatrix}
	,
\end{align}
where the proposal \eqref{eq:theoretical:peskunproposal} and acceptance probability \eqref{eq:theoretical:peskunaprob} enters into $p_{lm}$. Hence, we can estimate the \textit{acceptance rate} (the probability of a jump) by
\begin{align}
	P_{\text{jump}} = \sum_{l=1}^L \pi_l ( 1- p_{ll} ),
	\label{eq:theory:acceptrateestimate}
\end{align}
where the \textit{stationary probability} is calculated as
\begin{align*}
	\pi_l = \mathcal{N}(z_l; \sigma_{\Phi}, 1 ).
\end{align*}
Furthermore from the asymptotic results in \cite{Peskun1973} and \cite{KemenySnell1976}, we can estimate $\nu$ by
\begin{align}
	\widehat{\nu}(f,\pi,P)
	=
	\varphi^{\top}
	\big( 
	2 B Z - B - B A 
	\big)
	\varphi,
	\label{eq:theory:asympototicvarianceestimate}
\end{align}
where $\varphi=[\varphi(z_1),\varphi(z_2),\ldots,\varphi(z_L)]$ and $B=\mathsf{diag}(\pi_1,\pi_2,\ldots,\pi_L)$. Here, we introduce the \textit{fundamental matrix} by $Z=[ I - (P - A) ]^{-1}$ and the \textit{limiting matrix} by $A=\{a_{lm}\}$ with $a_{lm}=\pi_m	$.

\begin{figure}[t]
	\centering
	\includegraphics[width=\textwidth]{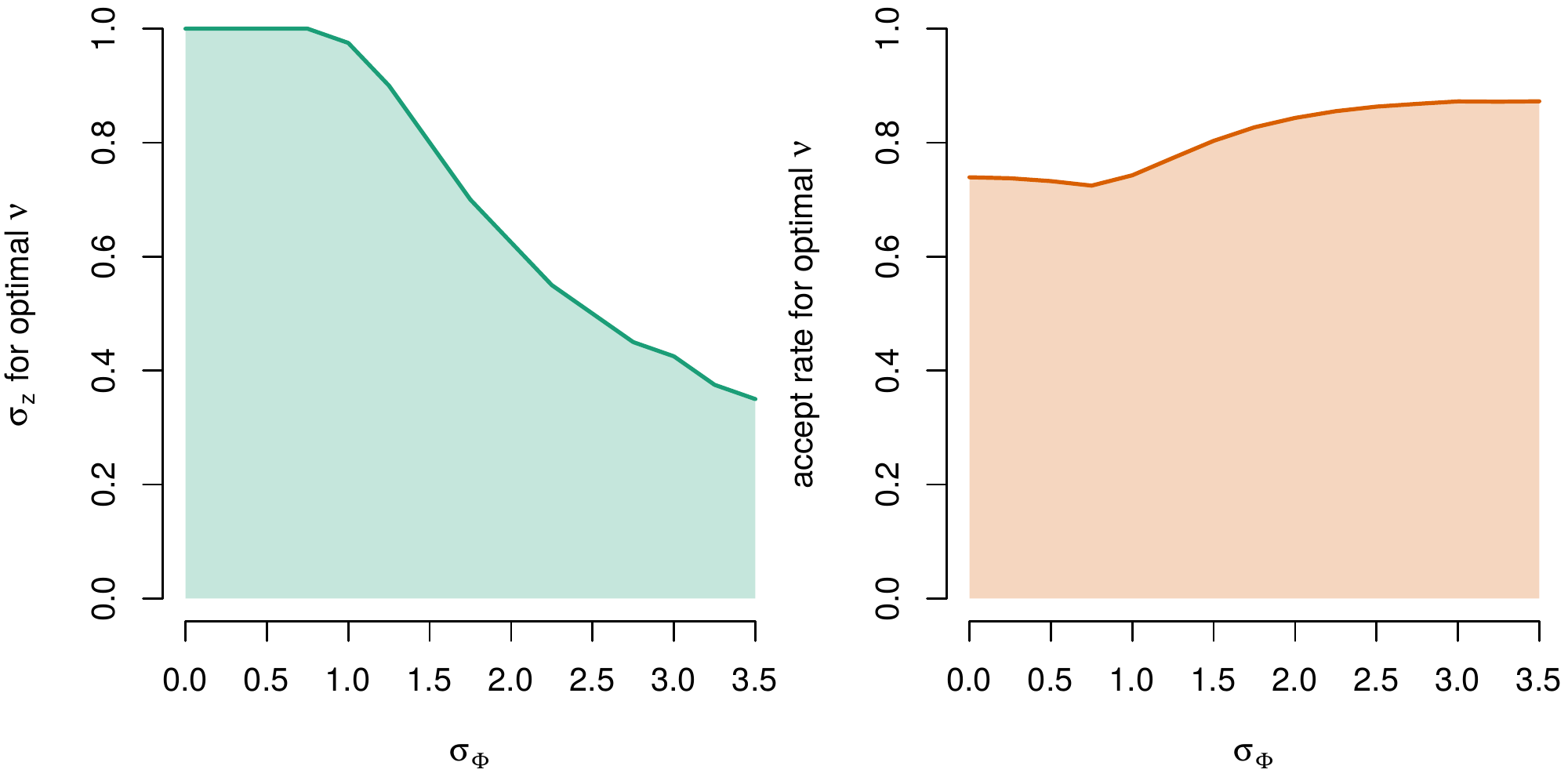}
	\caption{Optimal $\sigma_u$ (left) and acceptance rate (right) for minimising the $1/\widehat{\nu}(f,\pi,P)$.}
	\label{fig:peskun-analysis}
\end{figure}

\subsection{Tuning the CN proposal for the univariate case}
\label{sec:theory:results}
We make use of a modified version of the C-code provided by \cite{YangRodriguez2013} to implement the discretisation approach. We set $z_{\min}=-4$, $z_{\max}=\sigma_{\Phi}+4$ and $L=1,000$. Here, we use $\varphi(z)=z$ to optimise the mixing of the posterior mean. We vary $\sigma_{\Phi}$ in the target and $\sigma_z$ in the proposal in $\{0,0.25,\ldots,3.5\}$ and $\{0.05, 0.075, \ldots, 1\}$, respectively. For each $\sigma_{\Phi}$, we find the $\sigma_z$ that maximises $P_{\text{jump}}$ computed by \eqref{eq:theory:acceptrateestimate} and $1/\widehat{\nu}(f,\pi,P)$ calculated by \eqref{eq:theory:asympototicvarianceestimate}. In Figure~\ref{fig:peskun-analysis}, we present the resulting $\sigma_z$ is as a function of $\sigma_{\Phi}$. We note that for $\sigma_{\Theta} \in [1.0,1.8]$ as recommended by e.g.\ \cite{DoucetPittKohn2015}, we should have $\sigma_z \in [0.95,0.7]$ to obtain the optimal mixing and this corresponds to an acceptance rate of around $75 \%$ to $85 \%$ in the Markov chain for $u$.

Note that these results hold for the case when $u$ can be described by the univariate Gaussian random variable $z$. We return to numerically investigate how to optimal value of $\sigma_z$ in the CN proposal for the univariate random variable $z$ relates to the corresponding value $\sigma_u$ used in the CN proposal for the multivariate random variable $u$ in Section~\ref{sec:results}.

\section{Numerical illustrations}
\label{sec:results}
In this section, we investigate the numerical properties of the proposed alterations to the pmMH algorithm. We are especially interested in how the mixing is affected by making use of the CN proposal instead of an independent proposal for $u$. For this end with compare the mixing in two different models: (i) a Gaussian IID model with synthetic data and (ii) a nonlinear state space model with real-world data. The implementation details are presented in Appendix~\ref{app:impdetails}, where we also describe how to estimate the value of the target distribution (log-likelihood) for each model.

 We quantify the mixing by estimating the IACT using the empirical autocorrelation function. The estimate is computed by
\begin{align}
	\widehat{\textsf{IF}}(\theta_{K_b:K}) = 1 + 2 \sum_{\tau=1}^{100} \widehat{\rho}_{\tau} (\theta_{K_b:K}),
	\label{eq:results:ifdef}
\end{align}
where $\widehat{\rho}_{\tau}(\theta_{K_b:K})$ denotes the empirical lag-$\tau$ autocorrelation of $\theta_{K_b:K}$, and $K_b$ denotes the \textit{burn-in} time. 

\subsection{Gaussian IID model}
\label{sec:results:iid}
Consider the independent and identically (IID) distributed Gaussian model given by
\begin{align}
	x_0     \sim \delta_0, \qquad
	x_t     \sim \mathcal{N} \Big( x_t; \mu, \sigma_v^2 \Big), \qquad
	y_t|x_t \sim \mathcal{N} \Big( y_t; x_t, \sigma_e^2 \Big),
	\label{eq:results:iid}
\end{align}
with parameters $\theta=\{\mu,\sigma_v,\sigma_e\}$ and where $\delta_{x'}$ denotes the Dirac measure placed at $x=x'$. We generate a realisation with $T=10$ observations using the parameters $\theta^{\star}=\{0.5,0.3,0.1\}$.

\begin{figure}[t]
	\centering
	\includegraphics[width=\textwidth]{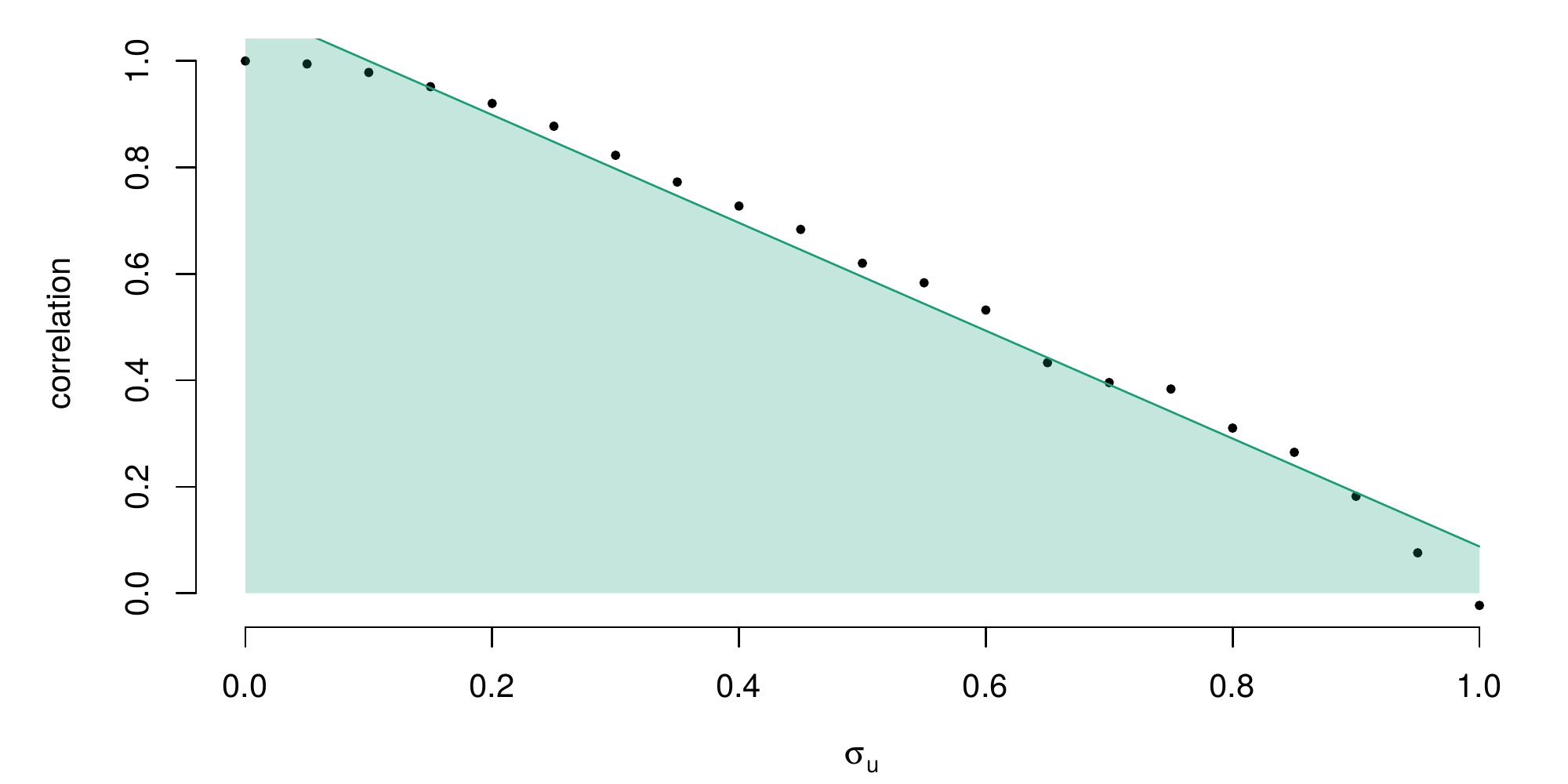}
	\caption{The estimated correlation (dots) in $\widehat{p}_{\theta}(y;u)$ for the Gaussian IID model \eqref{eq:results:iid} as a function of $\sigma_u$ and a linear regression (green). The linear approximation provides a reasonable approximation for much of the central part of the interval.}
	\label{fig:example1-correlation-versus-sigmau}
\end{figure}

We begin by investigating how the correlation of the log-likelihood estimated using importance sampling (see Appendix~\ref{app:impdetails:iid}) depends on $\sigma_u$. In Figure~\ref{fig:example1-correlation-versus-sigmau}, we present the correlation between two consecutive estimates of the log-likelihood (keeping $\mu$ fixed) when $\sigma_u \in \{0,0.05,\ldots,1.0\}$. We note that the correlation is almost linear when $\sigma_u > 0.2$ and can be described as $\mathsf{corr}( \widehat{p}_{\theta}(y;u), \widehat{p}_{\theta}(y;u') ) = 1 - \sigma_u$. We also estimate the standard deviation in the log-likelihood and obtain $3.5$. Hence, we conclude that this implies from Figure~\ref{fig:peskun-analysis} that the optimal correlation in the log-likelihood estimator is approximately $0.5$.

\begin{figure}[p]
	\centering
	\includegraphics[width=\textwidth]{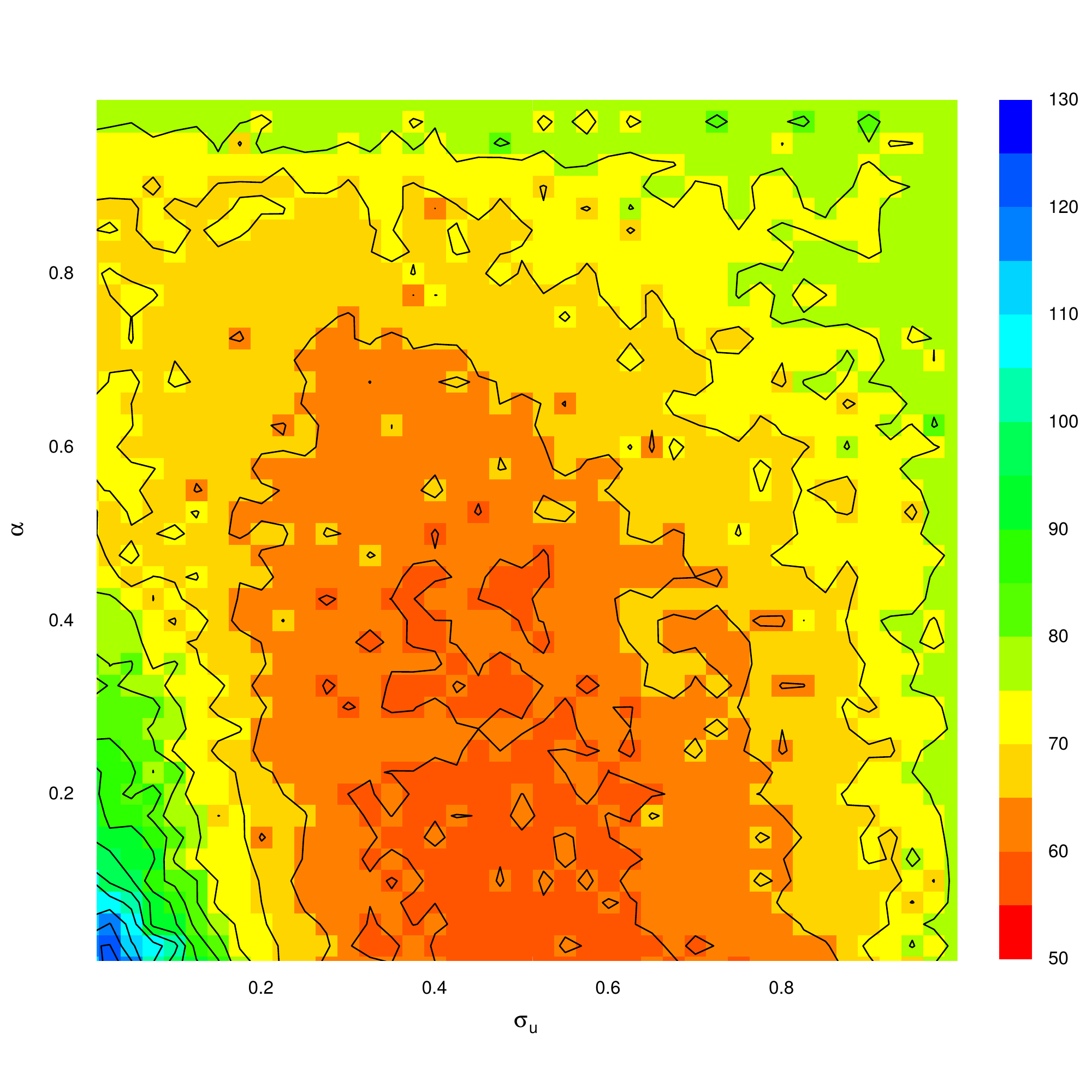}
	\caption{A heatmap of the IACT for the Gaussian IID model \eqref{eq:results:iid} when varying $\alpha$ and $\sigma_u$ in \eqref{eq:inducingcorrelation:extendedproposal}. The results presented are the median from $32$ independent Monte Carlo runs.}
	\label{fig:example1-iact-versus-sigmau-alpha}
\end{figure}

We proceed by fixing $\sigma_v$ and $\sigma_e$ to their true values and would like to infer the parameter posterior of $\mu$ given the data using Algorithm~\ref{alg:pmmh}. Here, the potential function corresponds to the log-likelihood estimator for the importance sampler as discussed in Appendix~\ref{app:impdetails:iid}. The aim is to compute the IACT over a grid of $\sigma_v$ and compare with the optimal theoretical results from Section~\ref{sec:theory}. We would also like to compare the proposed changes to pmMH with the suggestions discussed by \cite{LeeHolmes2010}. For this end, we consider a generalisation of the CN proposal \eqref{eq:inducingcorrelation:proposal} in which we introduce so-called \textit{global moves}. The mixture of local and global moves is a popular approach in e.g.\ computer graphics \citep{KelemenSzirmayAntalCsonka2002} to promote exploration of the entire target space. In our setup, we can introduce such a mixture into $q_{u}(u'|u_{k-1})$ and obtain the mixture proposal given by
\begin{align}
	\bar{q}_{\theta,u} \big( \{\theta',u\}|\{\theta,u\} \big)
	=
	q_{\theta} \big( \theta' | \{\theta,u\} \big)
	\Big[
	\alpha \,
	\mathcal{N} \big( u'; 0, \mathbf{I}_{N_u} \big)
	+
	( 1 - \alpha ) \,
	q_{u} \big( u'|u_{k-1} \big)
	\Big]
	\label{eq:inducingcorrelation:extendedproposal}
\end{align}
where we make use of the original proposals for $\theta$ and $u$ from \eqref{eq:inducingcorrelation:proposal}. Here, $\alpha \in [0,1]$ denotes the probability of a global move for $u$, where we recover the proposal in \eqref{eq:inducingcorrelation:proposal} by $\alpha=0$ and the proposal discussed by \cite{LeeHolmes2010} by $\sigma_u=0$ and $\alpha \neq 0$. We recover the standard pmMH proposal for $u$ when $\alpha=1$ and/or $\sigma_u=1$.

In Figure~\ref{fig:example1-iact-versus-sigmau-alpha}, we present a heatmap of the median IACT using \eqref{eq:inducingcorrelation:extendedproposal} when varying $\alpha,\sigma_u \in \{0,0.025,\ldots,1.0\}$. The smallest IACT values are obtain by using $\sigma_u \in [0.4, 0.6]$ with a small (or zero) value of $\alpha$. This range of $\sigma_u$ corresponds well with the results in Figure~\ref{fig:peskun-analysis} with $\sigma_{\Phi}=3.5$, which results in the optimal $\sigma_z=0.4$. Furthermore, we note that using $\sigma_u=0$ results in worse performance than if $\sigma_u > 0$ independently of the value of $\alpha$. Hence, we conclude that there is some benefit of using our proposed modification compared with the approach discussed by \cite{LeeHolmes2010} in this specific example. That is, using local moves seem to be more beneficial than global moves but there could be some merit to consider a mixture proposal in any case. However, we set $\alpha=0$ in the remainder of this section to simplify the tuning and make the conclusions more precise regarding the choice of $\sigma_u$.

\subsection{Stochastic volatility model with leverage}
\label{sec:results:sv}

Consider the problem of modelling the volatility in the daily closing prices of the NASDAQ OMXS30 index, i.e.\ a weighted average of the $30$ most traded stocks at the Stockholm stock exchange. We extract the $T=747$ log-returns using Quandl\footnote{The data is available for download from: \url{https://www.quandl.com/data/NASDAQOMX/OMXS30}.} for the period January 2, 2011 and January 2, 2014. To model the underlying volatility, we make use of a stochastic volatility model with leverage given by
\begin{align}
	x_0 \sim \mathcal{N} \Bigg( x_0; \mu, \frac{\sigma_v^2}{ \displaystyle \big( 1-\phi^2 \big)^2} \Bigg), \qquad
	\begin{bmatrix}	x_{t+1} \\ y_{t} \end{bmatrix} \Bigg| x_t
	\sim
	\mathcal{N}
	\left(
	\begin{bmatrix}	x_{t+1} \\ y_{t} \end{bmatrix};
	\begin{bmatrix}	\mu + \phi( x_t - \mu ) \\ 0 \end{bmatrix},
	\begin{bmatrix}	\sigma_v^2 & \rho \\ \rho & \exp(x_t) \end{bmatrix} 
	\right),
	\label{eq:results:svmodel}	
\end{align}
with parameters $\theta=\{\mu,\phi,\sigma_v,\rho\}$. In this model, we would like to infer the parameter posterior of $\theta$ given the data using Algorithm~\ref{alg:pmmh}. Here, the potential function corresponds to the log-likelihood estimator using the particle filter as discussed in Appendix~\ref{app:impdetails:sv}. The aim is to again compute the IACT and compare with the optimal theoretical results from Section~\ref{sec:theory}.

\begin{figure}[p]
	\centering
	\includegraphics[width=\textwidth]{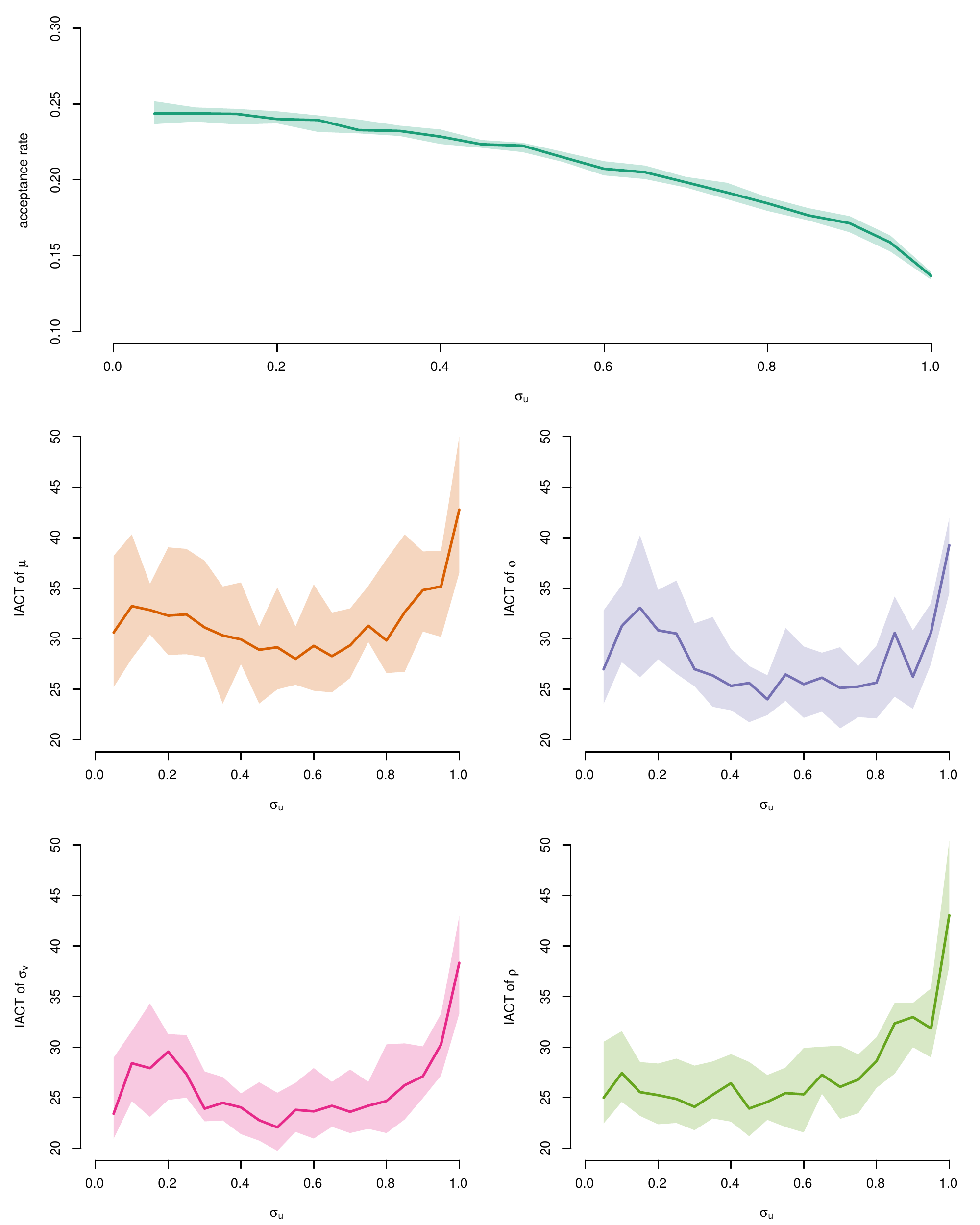}
	\caption{The acceptance probability (upper) and the resulting IACT (middle and lower) for $\mu$ (orange), $\phi$ (purple), $\sigma_v$ (magenta) and $\rho$ (light green) when varying $\sigma_u$. The results presented are the median (line) and first and third quantiles (shaded area) from $32$ independent Monte Carlo runs.}
	\label{fig:example3-iact-versus-sigmau}
\end{figure}

\begin{figure}[p]
	\centering
	\includegraphics[width=\textwidth]{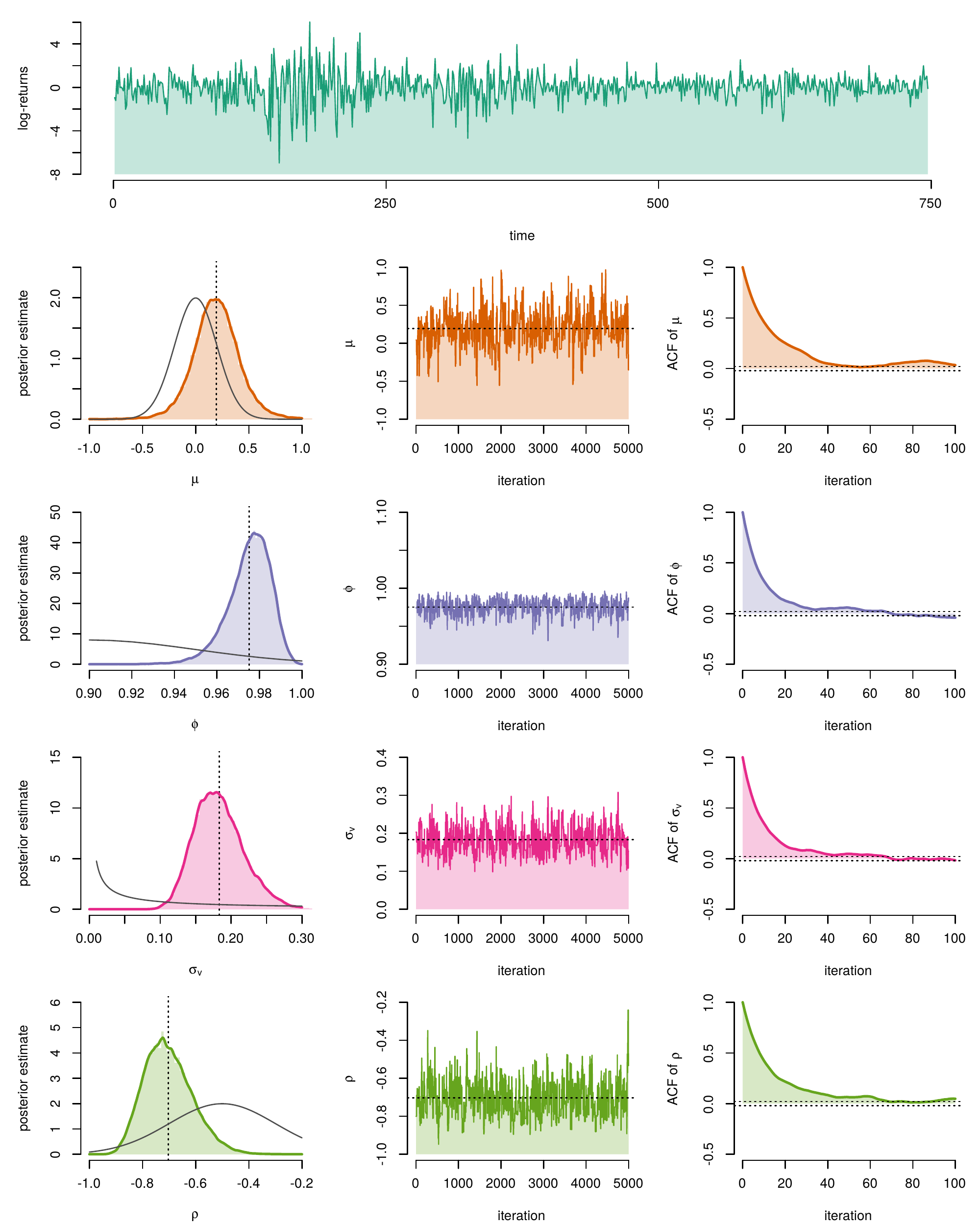}
	\caption{Upper: the closing log-returns for the NASDAQ OMXS30 index between January 2, 2011 and January 2, 2014. Lower: the parameter trace (left), autocorrelation function (center) and resulting posterior estimate (right) for the SV model \eqref{eq:results:svmodel} for $\mu$ (orange), $\phi$ (purple), $\sigma_v$ (magenta) and $\rho$ (light green). We make use of $\sigma_u=0.55$ in the CN proposal \eqref{eq:inducingcorrelation:proposal} and the histograms are computed using the output from $32$ independent Monte Carlo runs.}
	\label{fig:example3-diagonstic-posteriors}
\end{figure}

In Figure~\ref{fig:example3-iact-versus-sigmau}, we present the median acceptance rate and IACT for the four parameters in the model when varying $\sigma_u \in \{0.05, 0.10, \ldots, 1.00\}$. The maximum IACT (over the four variables) is minimised when $\sigma_u=0.55$. However, the variation in the IACT is significant and a range between $0.4$ and $0.8$ seems as suitable choices for $\sigma_u$. The standard deviation of the log-likelihood estimator (around the estimated posterior mean) is $1.2$, which by Section~\ref{sec:theory} would imply an optimal $\sigma_z$ of around $0.9$. This is clearly not an appropriate choice for this model. This can be due to that the Gaussian assumption for $\Phi_{\theta}(u)$ is not fulfilled or that the standard deviation of the log-likelihood estimate varies with $\theta$.

From Figure~\ref{fig:example3-iact-versus-sigmau}, we conclude that using $\sigma_u < 1$ results in a decrease of the IACT comparing with using an independent proposal for $u$. The improvement in the mixing is about $1.5$ times. We present a specific case in Figure~\eqref{fig:example3-diagonstic-posteriors}, where we fix $\sigma_u=0.55$. We conclude that the chains are mixing well and the posterior estimates are reasonable with the estimated posterior mean $\theta=\{0.19,0.98,0.18,-0.70\}$ with standard deviations $\{0.22,0.01,0.04,0.09\}$.

\section{Conclusions and future work}
\label{sec:conclusions}
The numerical illustrations in Section~\ref{sec:results} indicate that we can obtain improvements in the mixing of the pmMH algorithm by introducing correlation in $u$. In this paper, we consider using a CN proposal for introducing the correlation and it seems that selecting $\sigma_u=0.5$ offers good performance in the settings that we have considered. Hence, implementing the proposed alterations to the pmMH algorithm only requires changing a small number of lines of code and hopefully does not introduce any additional tuning parameters for the user. We are currently working on an adaptive method to further alleviate the work of tuning $\sigma_u$ for the user. The main benefit of our proposed modifications of the pmMH algorithm is that we can decrease the number of particles $N_u$ and obtain better mixing at the same time. Hence, this results in a significant decrease in the computational cost for Bayesian inference based on the pseudo-marginal framework.

Some important additional future work is to extended the theoretical analysis in Section~\ref{sec:theory} to a more realistic setting, where the influence of $\theta$ also is taking into the account. Also it would be interesting to make use of Hilbert curve resampling proposed by \cite{GerberChopin2015} or some tree methods discussed by \cite{Lee2008} in the particle filter. This would allow for inference in state space models with a multivariate state process. 

It is also possible to make use of $u$ to construct estimates of the gradient and Hessian of the log-posterior. This information can be included into the proposal for $\theta$ as discussed by \cite{DahlinLindstenSchon2015a,DahlinLindstenSchon2015c} to further improve the mixing in the Markov chain. The gradient information can also be used to reduce the variance in a post-processing step as discussed by \cite{MiraSolgiImparato2013}. 

The proposed alterations of the pmMH algorithm can also be useful in models where the likelihood is intractable as discussed by e.g.\ \cite{Jasra2015} and \cite{DahlinLindstenSchon2015c}. In this class of models, approximate Bayesian computations (ABC; \citealp{MarinPudloRobertRyder2012}  can be used to approximate the log-likelihood using importance sampling and particle filtering. However, in practice these estimates suffer from a large variance with results in bad mixing for the Markov chain. This approach can possible result in a large decrease in the computational cost for using the pmMH algorithm for inference in models with intractable likelihoods.

\section*{Acknowledgements}
This work was supported by: the Swedish Foundation for Strategic Research (SSF) through grant IIS11-0081, the project Probabilistic modeling of dynamical systems (Contract number: 621-2013-5524) funded by the Swedish Research Council and CADICS, a Linnaeus Center also funded by the Swedish Research Council. The simulations were performed on resources provided by the Swedish National Infrastructure for Computing (SNIC)  at Link\"{o}ping University, Sweden.

\clearpage
\bibliographystyle{plainnat}
\bibliography{dahlin}

\begin{thebibliography}{41}
\providecommand{\natexlab}[1]{#1}
\providecommand{\url}[1]{\texttt{#1}}
\expandafter\ifx\csname urlstyle\endcsname\relax
  \providecommand{\doi}[1]{doi: #1}\else
  \providecommand{\doi}{doi: \begingroup \urlstyle{rm}\Url}\fi

\bibitem[Andrieu and Roberts(2009)]{AndrieuRoberts2009}
C.~Andrieu and G.~O. Roberts.
\newblock {The pseudo-marginal approach for efficient Monte Carlo
  computations}.
\newblock \emph{The Annals of Statistics}, 37\penalty0 (2):\penalty0 697--725,
  2009.

\bibitem[Andrieu et~al.(2010)Andrieu, Doucet, and
  Holenstein]{AndrieuDoucetHolenstein2010}
C.~Andrieu, A.~Doucet, and R.~Holenstein.
\newblock {Particle Markov chain Monte Carlo methods}.
\newblock \emph{Journal of the Royal Statistical Society: Series B (Statistical
  Methodology)}, 72\penalty0 (3):\penalty0 269--342, 2010.

\bibitem[Andrieu et~al.(2012)Andrieu, Doucet, and Lee]{AndrieuDoucetLee2012}
C.~Andrieu, A.~Doucet, and A.~Lee.
\newblock {Discussion on constructing summary statistics for approximate
  Bayesian computation}.
\newblock \emph{Journal of the Royal Statistical Society: Series B (Statistical
  Methodology)}, 72\penalty0 (3):\penalty0 451--452, 2012.

\bibitem[Beskos et~al.(2008)Beskos, Roberts, Stuart, and
  Voss]{BeskosRobertsStuartVoss2008}
A.~Beskos, G.~Roberts, A.~Stuart, and J.~Voss.
\newblock {MCMC} methods for diffusion bridges.
\newblock \emph{Stochastics and Dynamics}, 8\penalty0 (03):\penalty0 319--350,
  2008.

\bibitem[Cotter et~al.(2013)Cotter, Roberts, Stuart, and
  White]{CotterRobertsStuartWhite2013}
S.~L. Cotter, G.~O. Roberts, A.~M. Stuart, and D.~White.
\newblock {MCMC methods for functions: modifying old algorithms to make them
  faster}.
\newblock \emph{Statistical Science}, 28\penalty0 (3):\penalty0 424--446, 2013.

\bibitem[Dahlin et~al.(2015{\natexlab{a}})Dahlin, Lindsten, and
  Sch\"{o}n]{DahlinLindstenSchon2015a}
J.~Dahlin, F.~Lindsten, and T.~B. Sch\"{o}n.
\newblock {Particle Metropolis-Hastings using gradient and Hessian
  information}.
\newblock \emph{Statistics and Computing}, 25\penalty0 (1):\penalty0 81--92,
  2015{\natexlab{a}}.

\bibitem[Dahlin et~al.(2015{\natexlab{b}})Dahlin, Lindsten, and
  Sch\"on]{DahlinLindstenSchon2015c}
J.~Dahlin, F.~Lindsten, and T.~B. Sch\"on.
\newblock {Quasi-Newton particle Metropolis-Hastings}.
\newblock In \emph{Proceedings of the 17th IFAC Symposium on System
  Identification (SYSID)}, Beijing, China, October 2015{\natexlab{b}}.

\bibitem[Dahlin et~al.(2015{\natexlab{c}})Dahlin, Villani, and
  Sch\"{o}n]{DahlinVillaniSchon2015}
J.~Dahlin, M.~Villani, and T.~B. Sch\"{o}n.
\newblock {Efficient approximate Bayesian inference for models with intractable
  likelihoods}.
\newblock \emph{Pre-print}, 2015{\natexlab{c}}.
\newblock arXiv:1506.06975v1.

\bibitem[Del~Moral et~al.(2006)Del~Moral, Doucet, and
  Jasra]{DelMoralDoucetJasra2006}
P.~Del~Moral, A.~Doucet, and A.~Jasra.
\newblock Sequential {M}onte {C}arlo samplers.
\newblock \emph{Journal of the Royal Statistical Society: Series {B}
  (Statistical Methodology)}, 68\penalty0 (3):\penalty0 411--436, 2006.

\bibitem[Deligiannidis et~al.(2015)Deligiannidis, Doucet, Pitt, and
  Kohn]{DeligiannidisDoucetPittKohn2015}
G.~Deligiannidis, A.~Doucet, M.~K. Pitt, and R.~Kohn.
\newblock {The Correlated Pseudo-Marginal Method}.
\newblock \emph{Pre-print}, 2015.
\newblock arXiv:1511.04992v1.

\bibitem[Doucet and Johansen(2011)]{DoucetJohansen2011}
A.~Doucet and A.~Johansen.
\newblock A tutorial on particle filtering and smoothing: Fifteen years later.
\newblock In D.~Crisan and B.~Rozovsky, editors, \emph{The Oxford Handbook of
  Nonlinear Filtering}. Oxford University Press, 2011.

\bibitem[Doucet et~al.(2015)Doucet, Pitt, Deligiannidis, and
  Kohn]{DoucetPittKohn2015}
A.~Doucet, M.~K. Pitt, G.~Deligiannidis, and R.~Kohn.
\newblock Efficient implementation of {M}arkov chain {M}onte {C}arlo when using
  an unbiased likelihood estimator.
\newblock \emph{Biometrika}, 102\penalty0 (2):\penalty0 295--313, 2015.

\bibitem[Fearnhead and Meligkotsidou(2015)]{FearnheadMeligkotsidou2015}
P.~Fearnhead and L.~Meligkotsidou.
\newblock {Augmentation schemes for particle MCMC}.
\newblock \emph{Statistics and Computing (in press)}, pages 1--14, 2015.

\bibitem[Flury and Shephard(2011)]{FluryShephard2011}
T.~Flury and N.~Shephard.
\newblock Bayesian inference based only on simulated likelihood: particle
  filter analysis of dynamic economic models.
\newblock \emph{Econometric Theory}, 27\penalty0 (5):\penalty0 933--956, 2011.

\bibitem[Gelman et~al.(1996)Gelman, Roberts, and Gilks]{GelmanRobertsGilks1996}
A.~Gelman, G.~Roberts, and W.~Gilks.
\newblock {Efficient Metropolis jumping rules}.
\newblock \emph{Bayesian statistics}, 5:\penalty0 599--607, 1996.

\bibitem[Gerber and Chopin(2015)]{GerberChopin2015}
M.~Gerber and N.~Chopin.
\newblock Sequential quasi {M}onte {C}arlo.
\newblock \emph{Journal of the Royal Statistical Society: Series {B}
  (Statistical Methodology)}, 77\penalty0 (3):\penalty0 509--579, 2015.

\bibitem[Hachisuka et~al.(2014)Hachisuka, Kaplanyan, and
  Dachsbacher]{HachisukaKaplanyanDachsbacher2014}
T.~Hachisuka, A.~S. Kaplanyan, and C.~Dachsbacher.
\newblock Multiplexed {M}etropolis light transport.
\newblock \emph{ACM Transactions on Graphics (TOG)}, 33\penalty0 (4):\penalty0
  100, 2014.

\bibitem[Hairer et~al.(2014)Hairer, Stuart, and
  Vollmer]{HairerStuartVollmer2014}
M.~Hairer, A.~M. Stuart, and S.~J. Vollmer.
\newblock {Spectral gaps for a Metropolis-Hastings algorithm in infinite
  dimensions}.
\newblock \emph{The Annals of Applied Probability}, 24\penalty0 (6):\penalty0
  2455--2490, 2014.

\bibitem[Hastings(1970)]{Hastings1970}
W.~K. Hastings.
\newblock {Monte Carlo sampling methods using Markov chains and their
  applications}.
\newblock \emph{Biometrika}, 57\penalty0 (1):\penalty0 97--109, 1970.

\bibitem[Jasra(2015)]{Jasra2015}
A.~Jasra.
\newblock {Approximate Bayesian Computation for a Class of Time Series Models}.
\newblock \emph{International Statistical Review}, 83\penalty0 (3):\penalty0
  405--435, 2015.

\bibitem[Kelemen et~al.(2002)Kelemen, Szirmay-Kalos, Antal, and
  Csonka]{KelemenSzirmayAntalCsonka2002}
C.~Kelemen, L.~Szirmay-Kalos, G.~Antal, and F.~Csonka.
\newblock A simple and robust mutation strategy for the {M}etropolis light
  transport algorithm.
\newblock \emph{Computer Graphics Forum}, 21\penalty0 (3):\penalty0 531--540,
  2002.

\bibitem[Kemeny and Snell(1976)]{KemenySnell1976}
J.~G. Kemeny and J.~L. Snell.
\newblock \emph{Finite Markov chains}.
\newblock Springer, New York, 1976.

\bibitem[Kloeden and Platen(1992)]{KloedenPlaten1992}
P.~E. Kloeden and E.~Platen.
\newblock \emph{Numerical solution of stochastic differential equations},
  volume~23.
\newblock Springer, 4 edition, 1992.

\bibitem[Kronander and Sch\"{o}n(2014)]{KronanderSchon2014}
J.~Kronander and T.~B. Sch\"{o}n.
\newblock {Robust auxiliary particle filters using multiple importance
  sampling}.
\newblock In \emph{Proceedings of the 2014 IEEE Statistical Signal Processing
  Workshop (SSP)}, Gold Coast, Australia, July 2014.

\bibitem[Lee(2008)]{Lee2008}
A.~Lee.
\newblock \emph{{Towards smooth particle filters for likelihood estimation with
  multivariate latent variables}}.
\newblock PhD thesis, University of British Columbia, 2008.

\bibitem[Lee and Holmes(2010)]{LeeHolmes2010}
A.~Lee and C.~Holmes.
\newblock {Discussion on Particle Markov chain Monte Carlo methods}.
\newblock \emph{Journal of the Royal Statistical Society: Series B (Statistical
  Methodology)}, 72\penalty0 (3):\penalty0 327--328, 2010.

\bibitem[Malik and Pitt(2011)]{MalikPitt2011}
S.~Malik and M.~K. Pitt.
\newblock Particle filters for continuous likelihood evaluation and
  maximisation.
\newblock \emph{Journal of Econometrics}, 165\penalty0 (2):\penalty0 190--209,
  2011.

\bibitem[Marin et~al.(2012)Marin, Pudlo, Robert, and
  Ryder]{MarinPudloRobertRyder2012}
J-M. Marin, P.~Pudlo, C.~P. Robert, and R.~J. Ryder.
\newblock {Approximate Bayesian computational methods}.
\newblock \emph{Statistics and Computing}, 22\penalty0 (6):\penalty0
  1167--1180, 2012.

\bibitem[Metropolis et~al.(1953)Metropolis, Rosenbluth, Rosenbluth, Teller, and
  Teller]{Metropolis1953}
N.~Metropolis, A.~W. Rosenbluth, M.~N. Rosenbluth, A.~H. Teller, and E.~Teller.
\newblock Equation of state calculations by fast computing machines.
\newblock \emph{{The Journal of Chemical Physics}}, 21\penalty0 (6):\penalty0
  1087--1092, 1953.

\bibitem[Meyn and Tweedie(2009)]{MeynTweedie2009}
S.~P. Meyn and R.~L. Tweedie.
\newblock \emph{Markov chains and stochastic stability}.
\newblock Cambridge University Press, 2009.

\bibitem[Mira et~al.(2013)Mira, Solgi, and Imparato]{MiraSolgiImparato2013}
A.~Mira, R.~Solgi, and D.~Imparato.
\newblock {Zero variance Markov chain Monte Carlo for Bayesian estimators}.
\newblock \emph{Statistics and Computing}, 23\penalty0 (5):\penalty0 653--662,
  2013.

\bibitem[Owen and Zhou(2000)]{OwenZhou2000}
A.~Owen and Y.~Zhou.
\newblock Safe and effective importance sampling.
\newblock \emph{Journal of the American Statistical Association}, 95\penalty0
  (449):\penalty0 135--143, 2000.

\bibitem[Peskun(1973)]{Peskun1973}
P.~H. Peskun.
\newblock {Optimum Monte-carlo sampling using Markov chains}.
\newblock \emph{Biometrika}, 60\penalty0 (3):\penalty0 607--612, 1973.

\bibitem[Pitt et~al.(2012)Pitt, Silva, Giordani, and
  Kohn]{PittSilvaGiordaniKohn2012}
M.~K. Pitt, R.~S. Silva, P.~Giordani, and R.~Kohn.
\newblock On some properties of {M}arkov chain {M}onte {C}arlo simulation
  methods based on the particle filter.
\newblock \emph{Journal of Econometrics}, 171\penalty0 (2):\penalty0 134--151,
  2012.

\bibitem[Robert and Casella(2004)]{RobertCasella2004}
C.~P. Robert and G.~Casella.
\newblock \emph{{Monte Carlo statistical methods}}.
\newblock Springer, 2 edition, 2004.

\bibitem[Ross(2012)]{Ross2012}
S.~M. Ross.
\newblock \emph{Simulation}.
\newblock Academic Press, 5 edition, 2012.

\bibitem[Sherlock et~al.(2015)Sherlock, Thiery, Roberts, and
  Rosenthal]{SherlockThieryRobetsRosenthal2015}
C.~Sherlock, A.~H. Thiery, G.~O. Roberts, and J.~S. Rosenthal.
\newblock {On the efficency of pseudo-marginal random walk Metropolis
  algorithms}.
\newblock \emph{The Annals of Statistics}, 43\penalty0 (1):\penalty0 238--275,
  2015.

\bibitem[Tierney(1994)]{Tierney1994}
L.~Tierney.
\newblock {M}arkov chains for exploring posterior distributions.
\newblock \emph{The Annals of Statistics}, 22\penalty0 (4):\penalty0
  1701--1728, 1994.

\bibitem[Tran et~al.(2014)Tran, Scharth, Pitt, and
  Kohn]{TranScharthPittKohn2014}
M.-N. Tran, M.~Scharth, M.~K. Pitt, and R.~Kohn.
\newblock {Importance Sampling Squared for Bayesian Inference in Latent
  Variable Models}.
\newblock \emph{Pre-print}, 2014.
\newblock arXiv:1309.3339v3.

\bibitem[Veach and Guibas(1997)]{VeachGuibas1997}
E.~Veach and L.~J. Guibas.
\newblock Metropolis light transport.
\newblock In \emph{Proceedings of the 24th International Conference on Computer
  Graphics and Interactive Techniques (SIGGRAPH)}, pages 65--76, Los Angeles,
  CA, USA, August 1997.

\bibitem[Yang and Rodr{\'\i}guez(2013)]{YangRodriguez2013}
Z.~Yang and Carlos~E Rodr{\'\i}guez.
\newblock {Searching for efficient Markov chain Monte Carlo proposal kernels}.
\newblock \emph{Proceedings of the National Academy of Sciences}, 110\penalty0
  (48):\penalty0 19307--19312, 2013.

\end{thebibliography}

\clearpage
\appendix
\section{Implementation details}
\label{app:impdetails}
In this appendix, we outline the implementation details of the numerical illustrations presented in Section~\ref{sec:results}. The implementations for estimating the log-likelihood are based on standard importance sampling and particle filtering. Extensive treatments of these methods are found in e.g.\ \cite{DoucetJohansen2011} and \cite{RobertCasella2004}.

\subsection{Gaussian IID model}
\label{app:impdetails:iid}
We make use of an importance sampler with $N=10$ to estimate the log-likelihood with the prior as the importance distribution. This results in that the log-likelihood can be estimated by
\begin{align}
	\log \widehat{p}_{\theta}(y;u)
	&=
	\sum_{t=1}^T
	\log
	\left[
	\sum_{i=1}^N
	w^{(i)}_t
	\right]
	- T \log N,
	\label{eq:app:llest}
\end{align}	
where the weights $w^{(i)}_t$ are generated using the procedure outlined in Algorithm~\ref{alg:is}.

\begin{algorithm}[b]
\caption{\textsf{Likelihood estimation using importance sampling with fixed random numbers}}
\textsc{Inputs:} $y$ (vector of $T$ observations), $N_u \in \mathbb{N}$ (no.\ samples) and $u \in \mathcal{U}$ ( $T \cdot N_u$ standard Gaussian random variables). \\
\textsc{Outputs:} $\widehat{p}_{\theta}(y;u)$ (est.\ of the likelihood). \\
\textsc{Note:} all operations are carried out over $i = 1, \ldots, N_u$.
\algrule[.4pt]
\begin{algorithmic}[1]
	\FOR{$t=1$ to $T$}
		\STATE Simulate from the proposal by
		\begin{align*}
		x_t^{(i)} = \mu + \sigma^2_v u^{(i)}_t,
		\end{align*} using random variables $u_{1:N_u,t}$.
		\STATE Calculate the weights by 
		\begin{align*}
		w^{(i)}_t = \mathcal{N}(y_t;x^{(i)}_t,\sigma_e^2).
		\end{align*}
	\ENDFOR
	\STATE Estimate $\widehat{p}_{\theta}(y;u)$ by \eqref{eq:app:llest}.
\end{algorithmic}
\label{alg:is}
\end{algorithm}

For pmMH, we use $K=10,000$ iterations (discarding the first $K_b=1,000$ as burn-in) in Algorithm~\ref{alg:pmmh} and initialise in the true parameter $\theta_0=0.5$. The proposal for $\theta$ is a standard Gaussian random walk proposal \eqref{eq:inducingcorrelation:proposal} with $\mu(\theta,u)=\theta$ and $\Sigma(\theta,u)=0.10^2$. Finally, we make use of the following prior
\begin{align*}
	p(\mu) \sim \mathcal{TN}_{(0,1)}(\mu;0,1),
\end{align*}
where $\mathcal{TN}_{(-1,1)}(\mu;0,1)$ denotes a standard Gaussian distribution truncated to the interval $(-1,1)$.

\subsection{Stochastic volatility model with leverage}
\label{app:impdetails:sv}
For any SSM, we make use of a bootstrap particle filter (bPF; \citealp{DoucetJohansen2011}) to estimate the log-likelihood. An SSM with latent states $x_{0:T}=\{x_t\}_{t=0}^T$ and observations $y_{1:T}$ is given by
\begin{align}
	x_0 \sim \mu_{\theta}(x_0), \qquad
	x_{t+1}|x_t \sim f_{\theta}(x_{t+1}|x_{t}), \qquad
	y_{t}  |x_t \sim g_{\theta}(y_{t  }|x_{t}),
\label{eq:SSMdef}
\end{align}
where $\theta \in \Theta \subseteq \mathbb{R}^p$ denotes the static unknown parameters. Here, we assume that it is possible to simulate from the distributions $\mu_{\theta}(x_0)$ and $f_{\theta}(x_{t+1}|x_{t})$ and evaluate $g_{\theta}(y_{t}|x_{t})$ point-wise. A bPF to estimate $\widehat{p}_{\theta}(y;u)$ is presented in Algorithm~\ref{alg:smc}. 

\begin{algorithm}[b]
\caption{\textsf{Likelihood estimation using particle filtering with fixed random numbers}}
\textsc{Inputs:} $y$ (vector of $T$ observations), an SSM \eqref{eq:SSMdef}, $N_u \in \mathbb{N}$ (no.\ particles) and $u \in \mathcal{U}$ ( $(T+1) \cdot (N_u+1)$ standard Gaussian random variables). \\
\textsc{Outputs:} $\widehat{p}_{\theta}(y;u)$ (est.\ of the likelihood). \\
\textsc{Note:} all operations are carried out over $i,j = 1, \ldots, N_u$.
\algrule[.4pt]
\begin{algorithmic}[1]
	\STATE Sample $x^{(i)}_0 \sim \mu_{\theta}(x_0)$ using $u_{2:(N+1),1}$.
	\STATE Set $W_0^{(i)}=N_u^{-1}$ as initial weights.
	\FOR{$t=1$ to $T$}
		\STATE Apply the inverse CDF approach to transform $u_{1,t+1}$ into a uniform random number $\bar{u}_{1,t+1}$.
		\STATE Apply systematic resampling with $\bar{u}_{1,t+1}$ to sample the ancestor index $a^{(i)}_t$ from a multinomial distribution with 
		\begin{align*}
		\mathbb{P} \big( a^{(i)}_t = j \big) = W^{(j)}_{t-1}.
		\end{align*}
		\STATE Propagate the particles by sampling 
		\begin{align*}
		x_t^{(i)} \sim f_{\theta} \Big( x_{t}^{(i)} \big| x_{t-1}^{a^{(i)}_t} \Big),
		\end{align*} using random variables $u_{2:(N_u+1),t+1}$.
		\STATE Extend the trajectory by $x_{0:t}^{(i)} = \Big\{ x_{0:t-1}^{a_t^{(i)}}, x_{t}^{(i)} \Big\}$.
		\STATE Sort the particle trajectories according to the current state $x_t^{(i)}$.
		\STATE Calculate the particle weights by $w^{(i)}_t = g_{\theta} \big( y_{t}|x_{t}^{(i)} \big)$ which by normalisation (over $i$) gives $W^{(i)}_t$. 
	\ENDFOR
	\STATE Estimate $\widehat{p}_{\theta}(y;u)$ by \eqref{eq:app:llest}.
\end{algorithmic}
\label{alg:smc}
\end{algorithm}

Hence, the log-likelihood is computed using the same expression \eqref{eq:app:llest} as for importance sampling but the particles $x^{(i)}_t$ are instead generated by sequential importance sampling with resampling. Note that we are required to sort the particles after the propagation step, see the discussion about smooth particle filter by \cite{MalikPitt2011} for more details. We make use of the probability transform to generate the uniform random variables required for the systematic resampling step. Hence, $u$ is a $(N_u+1) \cdot (T+1)$-variate Gaussian random variable, where $u_{2:N+1,t}$ is used directly in the propagation step and $u_{1,t}$ is used in the resampling step after a transformation into a uniform random variable. Here, we make use of $N_u=50$ particles.

For pmMH, we use $K=10,000$ iterations (discarding the first $K_b=1,000$ as burn-in) in Algorithm~\ref{alg:pmmh} and initialise the parameters at $\theta_0=\{0.23,  0.98 ,  0.18, -0.72\}$ obtained using the approach discussed by \cite{DahlinVillaniSchon2015}. The proposal for $\theta$ is a standard Gaussian random walk proposal \eqref{eq:inducingcorrelation:proposal} with $\mu(\theta,u)=\theta$. Here, we make use of the rules of thumb by \citet{SherlockThieryRobetsRosenthal2015} to select the covariance function. We extract an estimate of the Hessian using the method by \cite{DahlinVillaniSchon2015} and set
\begin{align*}
	\Sigma(\theta,u) = \frac{2.562^2}{p}
	\cdot
	10^{-4}
	\cdot
	\begin{bmatrix}
		 384&        3&         -5&         -16 \\
         3&          1&         -3&         -2 \\
        -5&         -3&          12&         3 \\
        -16&        -2&          3&          65 \\
	\end{bmatrix}
	.
\end{align*}
Finally, we use the following prior distributions
\begin{alignat*}{4}
	&p(\mu)      &&\sim \mathcal{N}(\mu;0,2^2), \qquad
	&p(\phi)      &\sim \mathcal{TN}_{(-1,1)}(\phi;0.9,0.05^2), \\
	&p(\sigma_v) &&\sim \mathcal{G}(\sigma_v;2.0,0.05), \qquad
	&p(\rho)     &\sim \mathcal{N}(\rho;-0.5,0.2^2),
\end{alignat*}
\noindent where $\mathcal{G}(a,b)$ denotes the Gamma distribution with mean $a/b$.

\end{document}